\documentclass[10pt,conference]{IEEEtran}
\IEEEoverridecommandlockouts

\usepackage{packages}
\begin{document}

\title{Do Developers Adopt Green Architectural Tactics for ML-Enabled Systems? A Mining Software Repository Study}
\author{\IEEEauthorblockN{Vincenzo De Martino}
\IEEEauthorblockA{Software Engineering (SeSa) Lab\\
University of Salerno, Italy\\
vdemartino@unisa.it}
\and
\IEEEauthorblockN{Silverio Martínez-Fernández}
\IEEEauthorblockA{Universitat Politècnica de Catalunya\\
Spain\\
silverio.martinez@upc.edu}
\and
\IEEEauthorblockN{Fabio Palomba}
\IEEEauthorblockA{Software Engineering (SeSa) Lab\\
University of Salerno, Italy\\
fpalomba@unisa.it}}

\maketitle
\begin{abstract}
As machine learning (ML) and artificial intelligence (AI) technologies become more widespread, concerns about their environmental impact are increasing due to the resource-intensive nature of training and inference processes. Green AI advocates for reducing computational demands while still maintaining accuracy. Although various strategies for creating sustainable ML systems have been identified, their real-world implementation is still underexplored.
This paper addresses this gap by studying 168 open-source ML projects on GitHub. It employs a novel large language model (LLM)-based mining mechanism to identify and analyze green strategies. The findings reveal the adoption of established tactics that offer significant environmental benefits. This provides practical insights for developers and paves the way for future automation of sustainable practices in ML systems.
\end{abstract}


\section{Introduction}
\label{sec:introduction}
Systems incorporating AI or ML, often referred to as AI-enabled or ML-enabled systems, are becoming increasingly common \cite{martinez2022software}. The growing use of these systems raises concerns about their sustainability, particularly in terms of energy consumption and CO\textsubscript{2} emissions, which have become increasingly pressing \cite{lacoste2019quantifying,duran2024identifying}. While the focus has traditionally been on maximizing accuracy, there is increasing recognition of the need to balance this with environmental impact, leading to the concept of Green AI \cite{schwartz2020green}. Green AI advocates for reducing computational demands while maintaining accuracy, emphasizing a responsible approach to AI development and deployment \cite{van2021sustainable}. In recent years, the software engineering (SE) community has made considerable efforts in promoting software sustainability, with specific attention now being paid to ML-enabled systems \cite{verdecchia2023systematic,tamburri2020sustainable,martinez2023towards}. This shift has identified various green tactics to reduce the environmental footprint of ML models. \cite{verdecchia2023systematic,10.1145/3639475.3640111}. For instance, J\"{a}rvenp\"{a}\"{a} et al. \cite{10.1145/3639475.3640111} 
cataloged 30 green architectural tactics for ML-enabled systems through a literature review and validated by experts. However,  the actual relevance of these tactics is not yet established. 

\steattentionboxa{\faWarning \hspace{0.05cm}
There is still a gap of knowledge on the extent to which green tactics are actually implemented in real-world contexts.}

Our objective seeks to determine the extent to which the green tactics cataloged by J\"{a}rvenp\"{a}\"{a} et al. \cite{10.1145/3639475.3640111} are adopted in the development of ML-enabled systems. Understanding this level of adoption is essential for bridging the gap between state-of-the-art research and the practices employed in real-world projects. Hence, we ask:

\rqquestion{To what extent are green architectural tactics for ML-enabled systems adopted in software projects?}

This paper addresses gaps in previous studies by conducting a mining software repository study to understand the adoption of green tactics in ML systems. Building on the catalog by J\"{a}rvenp\"{a}\"{a} et al.~\cite{10.1145/3639475.3640111}, the study analyzes 168 open-source ML projects on \textsc{GitHub} using a novel LLM-based mechanism to identify green tactics at the code level.
\begin{figure*}[ht]
    \centering
    \includegraphics[width=1\linewidth]{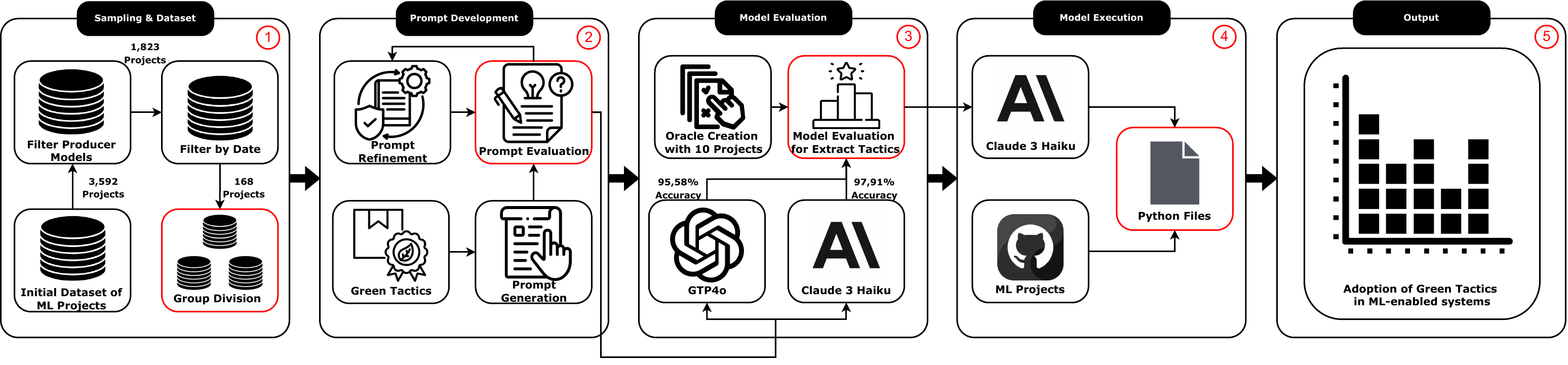}
    
    \caption{Overview of the research process.}
    \label{fig:overview}
\end{figure*}
\section{Related Work}
\label{sec:related}
Interest in sustainable and Green AI is growing, with the SE and AI communities publishing numerous studies to identify and catalog green practices that reduce energy consumption throughout the ML pipeline~\cite{verdecchia2023systematic,10.1145/3639475.3640111,natarajan2022theoretical,mittal2019survey}. However, most existing research focuses on identifying these practices rather than evaluating their adoption in real-world projects.


Salehi et al. \cite{salehi2023data} conducted a systematic review to classify data-centric approaches in Green AI, discussing practical applications, challenges, and future directions. Xu et al. \cite{xu2021survey} focused on Green Deep Learning, mapping approaches to reduce computational and energy costs in deep learning models while maintaining performance, using a taxonomy of lifecycle stages and artifacts. Building on this, Verdecchia et al. \cite{verdecchia2023systematic} expanded the scope to encompass the entire domain of Green AI, identifying a broader set of characteristics. J\"{a}rvenp\"{a}\"{a} et al. \cite{10.1145/3639475.3640111} later synthesized and validated these green practices with input from ML experts, classifying them into clusters, expanding on existing tactics, and identifying new ones.

To sum up, much work has been done to advance knowledge on what approaches can improve the fields of sustainable and green SE.  However, to the best of our knowledge, our study is the first to investigate the actual adoption of green architectural tactics in real software projects, marking the primary scientific contribution of our work. We specifically study the set of tactics defined in the literature, developing and evaluating an LLM-based tool to assess the adoption and frequency of these tactics in real-world projects. 

\begin{table*}[ht]
\caption{Average (AVG) and Median (Med) of each group of projects.}
\label{table:project_metrics_info}
\centering
\footnotesize
    \begin{tabular}{|c|c|c|c|c|c|c|c|c|c|c|c|c|c|}
    \rowcolor{black}
    \textcolor{white}{\textbf{Group}} & \textcolor{white}{\textbf{\#Projects}} & \multicolumn{2}{c|}{\textcolor{white}{\textbf{Age}}} & \multicolumn{2}{c|}{\textcolor{white}{\textbf{Stars}}} & \multicolumn{2}{c|}{\textcolor{white}{\textbf{Size}}} & \multicolumn{2}{c|}{\textcolor{white}{\textbf{Contributors}}} & \multicolumn{2}{c|}{\textcolor{white}{\textbf{Forks}}} & \multicolumn{2}{c|}{\textcolor{white}{\textbf{Commit}}} \\
    \rowcolor{black}
    \textcolor{white}{} & \textcolor{white}{} & \textcolor{white}{\textbf{Avg}} & \textcolor{white}{\textbf{Med}} & \textcolor{white}{\textbf{Avg}} & \textcolor{white}{\textbf{Med}} & \textcolor{white}{\textbf{Avg}} & \textcolor{white}{\textbf{Med}} & \textcolor{white}{\textbf{Avg}} & \textcolor{white}{\textbf{Med}} & \textcolor{white}{\textbf{Avg}} & \textcolor{white}{\textbf{Med}} & \textcolor{white}{\textbf{Avg}} & \textcolor{white}{\textbf{Med}} \\
    \hline
    Small & 60 & 6,3 & 5,67 & 1.101,68 & 255 & 3.466,75 & 2.629,50 & 9,35 & 4,5 & 244,57 & 65,5 & 532,63 & 291,5\\ 
    \hline
    \rowcolor{gray10} Medium & 59 & 6,67 & 6,63 & 2.543,78 & 669 & 46.739,27 & 42.446 & 15,08 & 13 & 412,85 & 148 & 1.114,46 & 691\\
    \hline
    Large & 49 & 6,98 & 6,6 & 6.703,76 & 563 & 435.995,41 & 248.857 & 19,8 & 21 & 1.681,33 & 154 & 4.257,47 & 1.506\\
    \hline
    \rowcolor{gray10} Total & 168 & 6,63 & 6,29 & 3.242,07 & 519,5 & 144.817,84 & 38.744,5 & 14,41 & 12 & 722,72 & 122,5 & 1.823,38 & 674\\
    \hline
    \end{tabular}
\end{table*}
\section{Research Method}
\label{sec:methodology}
This section discusses the methodology of our work; the main steps are overviewed in Figure \ref{fig:overview}. To report the results of our exploratory mining study, we adhere to the standards for repository mining as outlined in the \emph{``Repository Mining''} and \emph{``General Standard''} categories of the \textsl{ACM/SIGSOFT Empirical Standards}.\footnote{\url{https://github.com/acmsigsoft/EmpiricalStandards}.}

\subsection{Sampling and Dataset Description}
\label{sec:green_tactics}
We initially identified a candidate set of 3,592 ML projects coming from the combination of two popular datasets. The first dataset, proposed by Gonzalez \etal~\cite{gonzalez2020state} and later revised by Rzig \etal~\cite{rzig2022characterizing}, contains 3,020 ML projects. The second dataset, \textsc{NICHE}~\cite{widyasari2023niche}, includes 572 ML projects.  
The datasets include a variety of ML projects, from frameworks and tools to those creating or integrating ML models. However, not all were suitable for studying sustainable tactics in ML model lifecycles. Projects unrelated to ML model training were filtered out, narrowing the original 3,592 projects to a relevant subset aligned with the study's goals. Projects building and training ML models were identified by analyzing APIs for ML library imports and training method usage \cite{mluniverse}, using references such as the well-know AI \& ML Libraries list \cite{misiti2015awesome}. This step refined the dataset to 1,823 projects. Using the GitHub API \cite{GitHubREST}, only projects updated in 2024 were included, following prior research methodologies \cite{bernardo2024machine}. This further reduced the dataset to 168 projects, ensuring a focus on active projects.

The final set comprises 168 projects. \Cref{table:project_metrics_info} summarizes the characteristics of these projects, including metrics such as age, number of stars, and number of contributors. Additionally, we use the LOC to classify projects into three size groups: small (LOC $\leq$ 10.000), medium (10.000 $<$ LOC $\leq$ 100.000), and large (LOC $>$ 100,000). We use these groups to make comparisons in our analyses, which are in line with the methodology used by previous studies \cite{felidre2019continuous,bernardo2024machine}.

\subsection{Selection of Architectural Green Tactics} 
After selecting the ML projects for our study, we focused on identifying green architectural tactics using the comprehensive catalog by J\"{a}rvenp\"{a}\"{a} et al. ~\cite{10.1145/3639475.3640111}, which build upon the work of Verdecchia \etal~\cite{verdecchia2023systematic}.  This catalog, the most extensive resource available, organizes 30 green tactics into six clusters addressing various aspects of energy efficiency and sustainability across the ML pipeline. Unlike other studies focusing on domain-specific or isolated strategies~\cite{10.1145/3639475.3640111,natarajan2022theoretical,mittal2019survey}, it provides a holistic framework for analyzing green practices in ML systems.
From these candidate tactics, our study focuses on those pertaining to three specific clusters: \textsl{`Algorithm design'}, \textsl{`Model optimization'}, and \textsl{`Model training'}, encompassing a total of 15 tactics available in our appendix \cite{appendix}. These particular tactics are operational in nature, as they involve code-level and model-level adjustments that can be implemented using existing libraries or developed manually, depending on the developer's expertise. This flexibility ensures that the tactics are accessible to any development team, regardless of their resources, and can be effectively analyzed using the content available in \textsc{GitHub} repositories. While other tactics from the original study are undoubtedly valuable, they present significant challenges in terms of measurability. 

\subsection{Extraction of Architectural Green Tactics}
\label{sec:extractionTool}
Extracting architectural green tactics from software repositories may be challenging, as it requires discerning whether developers have implemented specific practices to support sustainability efforts. 

More specifically, we mined the latest snapshot of the repositories associated with the selected software projects, extracting all \textsl{Python} files, the files that may contain source code related to ML model creation, training, and deployment. To detect architectural green tactics, we applied LLMs directly to the extracted \textsl{Python} files. LLMs are highly effective at interpreting the source code, enabling them to identify subtle patterns and context-specific practices within these files \cite{lopes2024commit}. The advanced architecture and extensive training of LLMs enable them to capture complex code relationships and subtle patterns, making them effective for identifying context-specific green tactics in ML projects. By understanding source code context, e.g., recognizing optimized algorithms or model pruning for energy efficiency, LLMs can detect green tactics even without explicit documentation. We chose LLMs over other ML models for several reasons. Their architecture and diverse training enhance their ability to interpret code \cite{hou2023large}, and their capacity to process large volumes of code suits our study's scale. Moreover, LLMs can execute tasks from instructions without additional training data \cite{zhang2023prompting}, which is crucial given the lack of a labeled dataset for detecting green architectural tactics in software repositories.


To this aim, we devised an LLM-based mechanism aiming at detecting the application of green architectural tactics within software code. Our approach employs prompt engineering, where carefully crafted input prompts guide the LLM to generate outputs that accurately reflect the presence of tactics \cite{zhou2022large,radford2019language}. The prompt was structured into five components to achieve the study objective. First, the \textsl{`Objective'} defines the goal, instructing the LLM to analyze source code for sustainable practices, focusing on green architectural tactics for ML. Next, the \textsl{`Task Description'} outlines the task to detect 15 predefined sustainable tactics. The \textsl{`Additional Instructions'} guide the LLM on how to analyze the code and produce the output. In the \textsl{`Reporting Instructions'}, the LLM is directed to present findings clearly, including relevant code snippets, and explicitly note the absence of tactics when applicable. Lastly, the \textsl{`Examples'} section provides templates to ensure the results are formatted consistently across scenarios. We tested the prompt with GPT-4o (OpenAI) \cite{gptkey}, and Claude 3 Haiku (Anthropic) \cite{haikukey}, selected for their accuracy and ability to interpret complex unstructured data like source code \cite{chang2024survey}. The prompt was refined iteratively through collaborative sessions by the first two authors, using tests on a project subset to improve instructions and examples. The prompt's development process and iterations are detailed in the appendix \cite{appendix}.



\subsection{Model Validation}
To evaluate the feasibility of using LLMs to identify green architectural practices, we conducted an initial test with GPT-4o and Claude 3 Haiku on 10 projects (86 Python files) selected through random sampling~\cite{baltes2022sampling,thompson2012sampling}. The first author, leveraging expertise in sustainable SE, manually created an oracle to classify files based on the presence of sustainable tactics using a well-established catalog~\cite{10.1145/3639475.3640111}. This oracle served as the reference for assessing model accuracy.
GPT-4o achieved 95.58\% accuracy, offering precise but minimal contextual explanations. Claude 3 Haiku performed slightly better with 97.91\% accuracy, providing clear natural language explanations that enhanced understanding of its classifications. Despite occasionally omitting code snippets, Claude 3 Haiku's insights and cost efficiency (\$0.25/\$1.25 per million tokens vs. GPT-4o \$5/\$15) made it more suitable for analyzing larger datasets.
Ultimately, we selected Claude 3 Haiku for its higher accuracy and lower cost. To address its occasional omission of specific code snippets, we implemented a mechanism to track the repository and file where a tactic was detected, ensuring reliable localization even without explicit code output.

\subsection{Data Analysis}
\label{sec:extraction}
To ease the data analysis process, we developed a Python script ~\cite{appendix} that systematically traverses the files in each repository and records key information. For each project, it generates a \texttt{csv} file, with each row corresponding to an analyzed file. The \texttt{csv} file captures (1) the file name and (2) the output generated by Claude 3 Haiku. In particular, we implemented a script that processed Haiku's responses by segmenting the first part of the output---specifically the classification of existing tactics---into separate columns, with each column representing one of the tactics considered in the study. After collecting all responses, the data was aggregated into a single file that tracked the occurrence of each tactic. The script calculated the total frequency of each tactic across all projects and used this information to address the \textbf{RQ}, providing an overview of how green architectural tactics are adopted in the considered ML-enabled systems.

\subsection{Threats to validity}
\label{sec:threats}
This section highlights potential factors influencing the study’s conclusions and generalizability \cite{wohlin2012experimentation}. Errors from poorly designed prompts or misinterpreted tactic definitions were mitigated through iterative refinement and the use of well-established definitions from the literature \cite{10.1145/3639475.3640111}. Risks included randomness in LLM outputs and incomplete tactic extraction, which were addressed by refining prompts, conducting tests, and tracking repositories for verification. The study focuses on Python due to its prominence in ML, ensuring generalizability through dataset quality by using well-established sources like \textsc{NICHE}\cite{widyasari2023niche} and the dataset by Gonzalez et al.\cite{gonzalez2020state}, with preprocessing to exclude outdated or irrelevant projects. Sustainable tactics were selected based on J\"{a}rvenp\"{a}\"{a} et al.\cite{10.1145/3639475.3640111} catalog for their suitability for code-level analysis. Default LLM parameters were used to ensure consistency in evaluation. While based on a sizable sample, the findings cannot be fully generalized to all ML projects. A comparative evaluation of GPT-4o \cite{gptkey} and Claude 3 Haiku \cite{haikukey}, using a manually constructed oracle, was conducted to identify the reliable model for accuracy and tactic discovery.

\section{Analysis of the Results}
\label{sec:result}
\begin{figure}[h]
    \centering
    \includegraphics[width=1\linewidth]{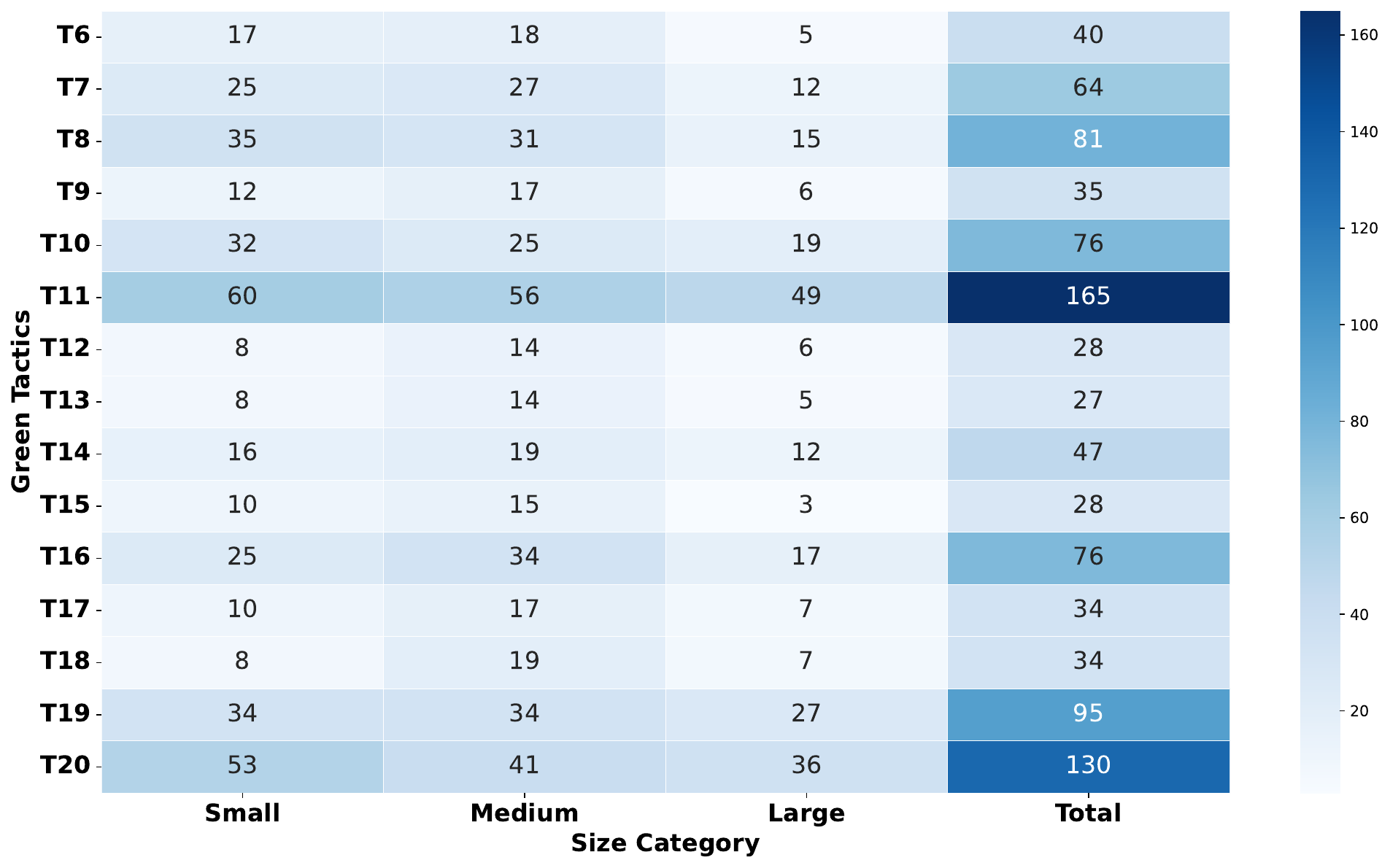}
    \caption{Frequency of Green Tactics in ML Projects.}
    \label{fig:frequency}
\end{figure}
Figure \ref{fig:frequency} shows the adoption frequency of green tactics across 168 ML projects, categorized by size: \textsl{`Small'}, \textsl{`Medium'}, \textsl{`Large'}, and the full set. \textsl{`Medium'} projects showed the highest overall adoption of tactics, with \textsl{`Small'} projects favoring tactics like \textsl{`Decrease model complexity'}, and \textsl{`Use dynamic parameter adaptation'}. \textsl{`Large'} projects demonstrated the lowest adoption rates, likely due to their complexity and a focus on stability and performance, which can hinder the integration of sustainable tactics \cite{gonzalez2009macro}. Across all projects, the most widely adopted tactic was \textsl{`Use of built-in library functions'} 165 (98.2\%), reflecting the reliance on optimized libraries and frameworks. \textsl{`Design for memory constraints'} 130 (77.4\%) was also prevalent, emphasizing efficiency in memory usage. Other frequently adopted tactics included  \textsl{`Use of checkpoints during training'} 95 (56.5\%), \textsl{`Reducing model complexity'} 81 (48.2\%), \textsl{`Use dynamic parameter adaptation'} 76 (45.2\%) and \textsl{`Consider transfer learning'} 76 (45.2\%). In contrast, tactics like \textsl{`Enhance model sparsity'} 47 (28\%) and \textsl{`Choose a lightweight algorithm alternative'} 40 (23.8\%) were less common. The least adopted tactics were \textsl{`Consider graph substitution'}, \textsl{Set energy consumption as a model constraint'}, and \textsl{`Consider energy aware pruning'} 27-28 (16.1\%-16.7\%). These low adoption rates may stem from their complexity or a lack of awareness among developers. This variation underscores the need for targeted education and tooling to facilitate the adoption of less-used but potentially impactful green architectural tactics.

\rqanswer{Green tactics proposed in the literature are widely adopted in ML projects. However, some tactics are less adopted, suggesting the need to raise awareness and give more support to the tools to facilitate their wider integration.}
\section{Discussion and Implications}
\label{sec:discussion}
Our study highlights several implications for the SE community and society to reduce the energy footprint. When comparing the adoption of green tactics identified in previous studies, particularly the catalog of 30 tactics by J\"{a}rvenp\"{a}\"{a} et al. \cite{10.1145/3639475.3640111}, we analyzed 168 open-source ML projects on GitHub using LLMs. Our analysis confirms that practitioners adopt many of these tactics. Specifically, tactics like \textsl{`Use of built-in library functions'} and \textsl{`Design for Memory Constraints'} have high adoption rates, suggesting that these tactics are well-supported and effectively integrated into current development processes. Conversely, other tactics such as \textsl{`Set Energy Consumption as a Model Constraint'} and \textsl{`Consider Graph Substitution'} exhibit lower adoption rates in our study. 
Although our analysis is based on open-source projects, the results have broader implications for different ML contexts. Many tactics are applicable to proprietary and enterprise-level systems, but some, such as energy-aware optimizations, may require adaptations for specific domains such as resource-constrained environments. Green tactics also hold potential for improving accessibility and affordability by reducing hardware and energy costs, but their adoption must be balanced against potential trade-offs in performance and usability, particularly in critical applications such as healthcare and autonomous systems. Adoption barriers, including organizational resistance, lack of awareness, and perceived costs, are particularly evident in larger projects where coordination and competing priorities complicate sustainability efforts. Addressing these barriers will require better education, improved tooling, and incentives for adopting green tactics. Engaging directly with developers through interviews or analyzing project repositories could uncover additional insights into these challenges and inform strategies for overcoming them.

\section{Conclusion and Future Work}
\label{sec:conclusion}
This paper examines the adoption of green tactics in ML-enabled systems through a study of 168 open-source ML projects and introduces a novel LLM-based mechanism for identifying green tactics in software repositories. Our contributions include an empirical analysis of green tactic adoption, a novel LLM tool to identify tactics, and a public replication package with study materials \cite{appendix}. 
Future work will explore the impact of green tactics on sustainability through longitudinal studies and the identification of new green tactics. It will also examine socio-technical factors, such as knowledge gaps and implementation challenges while promoting lesser-used tactics with enhanced documentation, tools, and case studies. These steps will help integrate green tactics more effectively into ML development, contributing to the creation of more sustainable systems.

\section*{Acknowledgment}
This work has been partially supported by the European Union - NextGenerationEU through the Italian Ministry of University and Research, Projects PRIN 2022 "QualAI: Continuous Quality Improvement of AI-based Systems" (grant n. 2022B3BP5S, CUP: H53D23003510006), and by the GAISSA research project (ref. TED2021-130923B-I00; MCIN/AEI/10.13039/501100011033).

\balance
\bibliographystyle{IEEEtran}
\bibliography{bib/bib}

\end{document}